\documentclass[a4paper,12pt]{article} 
\usepackage{amsfonts,amsmath,epsfig,subfigure,harvard} 
\begin{document} 
\numberwithin{equation}{section} 
\title{Elastic interface acoustic waves \\ in twinned crystals.} 

\author{Michel Destrade}
\date{2003} 
\maketitle 

\bigskip


\begin{abstract} 
A new type of Interface Acoustic Waves (IAW) is presented, 
for single-crystal orthotropic twins bonded symmetrically along a 
plane containing only one common crystallographic axis. 
The effective boundary conditions show that the waves are linearly 
polarized at the interface, either transversally or longitudinally. 
Then the secular equation is obtained in full analytical form using 
new relationships for the displacement-traction quadrivector at the 
interface. 
For Gallium Arsenide and for Silicon, it is found that the IAWs 
with transverse (resp. longitudinal) polarization at the interface are 
of the Stoneley (resp. leaky) type.

\end{abstract} 

\vspace{12pt}


\newpage

\section{Introduction} 

The possibilities of existence of Stoneley waves propagating at the 
interface of two welded half-spaces are quite restricted, especially 
when the semi-infinite bodies are made of \textit{dissimilar} 
crystals. 
For example, Owen (1964) famously studied 900 combinations of 
isotropic materials and found only 31 pairs supporting Stoneley waves. 
Similar conclusions are also reached when the materials are 
anisotropic.
In general, the existence of an interfacial Stoneley wave is highly 
sensitive to the differences in material parameters for each medium, 
in particular to the difference in shear wave speeds 
(Chadwick \& Currie, 1974). 

The situation is somewhat more favorable when the half-spaces are 
made of misoriented but \textit{identical} crystals. 
For instance, Stoneley waves were found to exist for any angle of 
misorientation and for propagation along the twist-angle bisectrix in 
the case of some hypothetical crystal by Lim \& Musgrave (1970), 
Copper  by Th\"ol\'en (1984), Gallium Arsenide by Barnett et al. 
(1985), or Silicon by Mozhaev et al. (1998); 
they also exist for 180 degrees domain boundaries of Barium 
Titanate or Quartz 
(Mozhaev \& Weihnacht, 1996, 1997), for propagation 
in any direction in the interface plane.
In these articles, the interface is normal to a crystallographic 
axis which is itself common to each half-space.

\begin{figure}
\centering
\mbox{\epsfig{figure=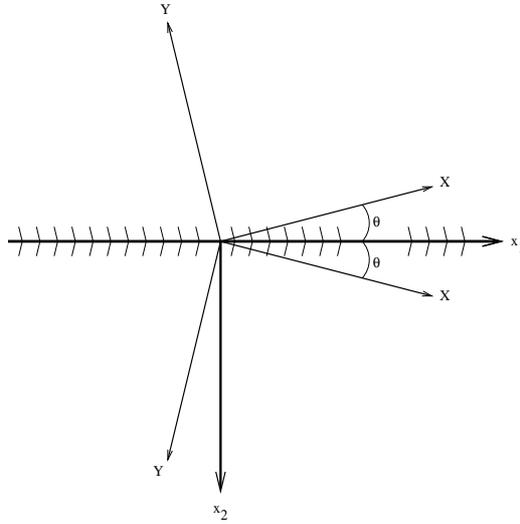, height=.5\textwidth, width=.5\textwidth}}
 \caption{Twinned crystal}
\end{figure}

Now consider the interface described in Figure 1, where the upper and 
lower half-spaces are made of the same crystal with at least rhombic 
symmetry and with misoriented crystallographic axes $X$ and $Y$ 
(represented on the Figure making an angle $\pm \theta$ with the 
interface and its normal) and common crystallographic axis $Z$ 
(normal to the plane of the Figure).
This geometry of chevron or ``herringbone'' pattern might be 
encountered in crystals subjected to a gliding twinning plane, or to 
the formation of conjugate kink bands through compression. 
Hussain and Ogden (2000) recently proposed a theoretical 
simulation of the plastic deformation associated with the twinning 
of crystals, with a possible exploitation in non-destructive testing 
of materials.
Another way of producing such twins would be the following. 
Consider an infinite crystal with orthotropic or higher symmetry; 
cut it in two halves so that the plane interface contains one 
crystallographic axis ($x_3$ say) 
and makes an angle $\theta$ with another crystallographic axis 
($x_1$ say), see Figure 2(a);
then rotate one half-space by 180 degrees about the normal to the 
interface, see Figure 2(b); finally rebond the half-spaces, see Figure 
2(c).
A thorough review article by G\"osele and Tong (1998) presents 
several experimental procedures of wafer bonding (also known as 
direct bonding or fusion bonding or ``gluing without glue'') and its 
numerous applications, not only for Silicon-based sensors and 
actuators but also in microsystem technologies, nonlinear optics, 
light-emitting diodes, etc. using Gallium Arsenide, Quartz, 
or Sapphire.

\begin{figure}
\centering
\mbox{\epsfig{figure=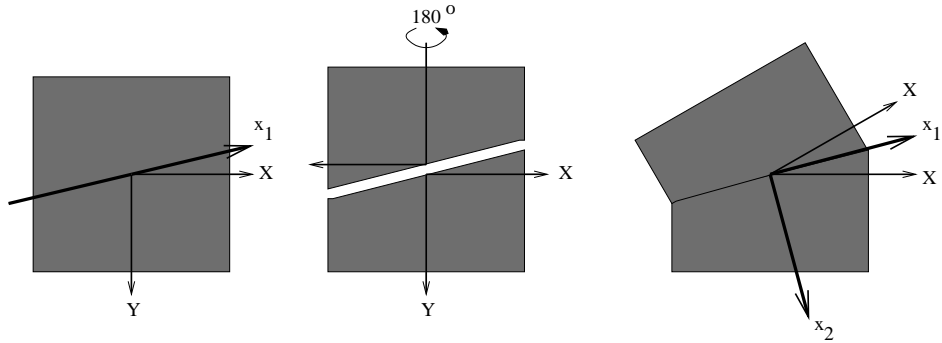, width=.9\textwidth}}
 \caption{Cutting (a), rotating (b), and bonding (c) 
of a rhombic crystal}
\end{figure}

When the boundary conditions take the symmetries of the problem into 
consideration then the interface wave is found to be linearly 
polarized at the interface, either longitudinally or transversally.
Once these effective boundary conditions are established in Section 2 
(and Appendix), two corresponding secular equations are derived 
explicitly in Section 3 as cubics in the squared wave speed. 
The results are illustrated graphically for Gallium Arsenide
and for Silicon. 
For these materials, it turns out that one of the secular equations 
corresponds to a non-physical supersonic leaky interface wave, and 
that the other corresponds to a subsonic interface acoustic wave 
having the characteristics of a Stoneley wave: 
decay with increasing distance from the interface, 
speed larger than that of the corresponding surface Rayleigh wave.

\section{Effective boundary conditions} 

Consider Stoneley waves traveling with speed $v$ and wave number $k$ 
at the interface of a bimaterial made of two perfectly bonded 
orthotropic media.
Both half-spaces are made of the same crystal (mass density $\rho$, 
non-zero reduced compliances $s'_{11}, s'_{22}, s'_{12}, s'_{44}, 
s'_{55}, s'_{66}$).
However, the normal to the interface $Ox_2$ and the direction of 
propagation $Ox_1$ are inclined at an angle $\theta$ to the 
crystallographic axes $Oy$ and $Ox$ of the lower ($x_2>0$) half-space 
and at an angle $-\theta$ for the upper ($x_2<0$) half-space. 

In the lower half-space, the strain-stress relation is 
$\epsilon_{ij} = s_{ik} \sigma_{ik}$ where the reduced compliances 
$s_{ik}$ are given by (see for instance Ting (2000) or 
Destrade (2003)), 
\begin{align} \label{s'rotated}
& s_{11} = s'_{11}\cos^4 \theta 
            + (2s'_{12} + s'_{66})\cos^2 \theta \sin^2 \theta 
             + s'_{22}\sin^4 \theta,
\nonumber \\
& s_{22} = s'_{22}\cos^4 \theta 
            + (2s'_{12} + s'_{66})\cos^2 \theta \sin^2 \theta
             + s'_{11}\sin^4 \theta,
\nonumber \\
& s_{12} = 
  s'_{12} 
    + (s'_{11}+s'_{22}-2s'_{12}-s'_{66})\cos^2 \theta \sin^2 \theta,
\nonumber \\
& s_{66} = 
s'_{66}
  + 4(s'_{11}+s'_{22}-2s'_{12}-s'_{66})\cos^2 \theta \sin^2 \theta.
\nonumber \\
& s_{16} = 
  [2s'_{22}\sin^2 \theta  - 2s'_{11}\cos^2 \theta
   + (2s'_{12} + s'_{66})(\cos^2 \theta - \sin^2 \theta)]
                                             \cos \theta \sin \theta,
\nonumber \\
& s_{26} = 
  [2s'_{22}\cos^2 \theta  - 2s'_{11}\sin^2 \theta
    - (2s'_{12} + s'_{66})(\cos^2 \theta - \sin^2 \theta)]
                                             \cos \theta \sin \theta.
\end{align}
When the mechanical displacement $\mathbf{u}(x_1,x_2,x_3,t)$ is 
modelled as a linear combination of the partial modes 
$e^{ik(x_1 + px_2 - vt)}\mathbf{U}$ (say), then the in-plane strain 
decouples from the anti-plane strain, and so does in-plane stress  
from anti-plane stress (Stroh, 1962);
also, $p$ is a root of the characteristic polynomial , 
\begin{equation} \label{quartic-p}
\omega_4 p^4 - 2\omega_3 p^3 + \omega_2 p^2 - 2 \omega_1 p 
   + \omega_0 = 0,
\end{equation}
where the (real) coefficients $\omega_i$ are given in terms of the 
reduced compliances and of $X = \rho v^2$ by (Destrade, 2003), 
\begin{align} \label{omegas}
& \omega_4 = s_{11}, \quad 
  \omega_3 = s_{16}, \nonumber \\
& \omega_2 = s_{66} + 2s_{12}
                - [s_{11}(s_{22} + s_{66}) - s_{12}^2 - s_{16}^2]X, 
\nonumber  \\
& \omega_1 = s_{26} +
                 [s_{16}(s_{22} - s_{12}) + s_{26}(s_{11} - s_{12})]X, 
\nonumber  \\
& \omega_0 = s_{22}
                - [s_{22}(s_{11} + s_{66}) - s_{12}^2 - s_{26}^2]X
                    +   \begin{vmatrix}
                         s_{11} & s_{12} & s_{16} \\  
                         s_{12} & s_{22} & s_{26} \\  
                         s_{16} & s_{26} & s_{66}
                        \end{vmatrix} X^2. 
\end{align} 

In the upper half-space, $\theta$ is changed to its opposite 
so that by \eqref{s'rotated}, $s_{11}$, $s_{22}$, $s_{12}$, $s_{66}$ 
remain unchanged whilst $s_{16}$ and $s_{26}$ change signs. 
Consequently, by  \eqref{omegas} the characteristic polynomial 
in the upper half-space is 
\begin{equation} \label{quartic+p}
\omega_4 p^4 + 2\omega_3 p^3 + \omega_2 p^2 + 2 \omega_1 p 
   + \omega_0 = 0.
\end{equation}
It follows that if $p$ is a root of the characteristic polynomial 
\eqref{quartic-p} for the lower half-space, then $-p$ is a root of the 
characteristic polynomial \eqref{quartic+p} for the upper half-space. 

As a consequence of these properties (change of sign across the 
interface for $s_{16}$, $s_{26}$, and for the $p$'s), the following 
effective boundary conditions apply at $x_2=0$ (see Appendix),
\begin{equation} \label{BC}
u_1 = \sigma_{22} = 0, 
\quad \text{or} \quad 
u_2 = \sigma_{12} = 0.
\end{equation}
Following Mozhaev et al. (1998), interface acoustic waves satisfying 
the first condition \eqref{BC}$_1$ are denoted IAW1, and those  
satisfying the second condition \eqref{BC}$_2$ are denoted IAW2.
The polarization of the waves at the interface is linear: 
transverse for IAW1 and longitudinal for IAW2. 
Note that Mozhaev and collaborators obtained similar effective 
boundary conditions for other types of twin boundaries 
(180$^\text{o}$ ferroelectric domain boundary in tetragonal 
single-crystals with crystallographic propagation direction 
(Mozhaev \& Weihnacht,  1995, 1996); 
Dauphin\'e twins in quartz with boundary coincident with the 
$YZ$ mirror plane and propagation along the $Y$ direction 
(Mozhaev \& Weihnacht, 1997); 
twisted single-crystal cubic wafers with basal plane boundary and 
propagation along the twist-angle bisectrix 
(Mozhaev \& Tokmakova, 1994; Mozhaev et al., 1998).

\section{Equations of motion and explicit secular equations} 

The equations of motion in the lower half-space are written as a 
first-order homogeneous differential system for the in-plane 
displacement-traction vector,
\begin{equation}  
\mbox{\boldmath $\xi$}(kx_2) = 
 [U_1(kx_2), U_2(kx_2), t_{12}(kx_2), t_{22}(kx_2)]^\text{T}, 
\end{equation}
where the $U_i$ and $t_{i2}$ are defined by
\begin{equation} 
u_i(x_1, x_2, x_3, t) = U_i(kx_2)e^{ik(x_1 - vt)}, 
\quad 
\sigma_{i2} (x_1, x_2, x_3, t) = ikt_{i2}(kx_2)e^{ik(x_1 - vt)}.
\end{equation}
Explicitly, the system is given by (Destrade, 2001), 
\begin{equation} \label{EqnMotion}
\mbox{\boldmath $\xi$}'= i\mathbf{N}\mbox{\boldmath $\xi$}, 
\quad 
\mathbf{N} =
 \begin{bmatrix} 
  -r_6 &  - 1  & n_{66}  &  n_{26}  \\
  -r_2 &    0  & n_{26}  &  n_{22}  \\
 X-\eta&    0  &  -r_6   &  -r_2    \\  
   0   &    X  &  - 1    &    0   
 \end{bmatrix}, 
\end{equation}
where $X=\rho v^2$ and (Ting, 2002)
\begin{equation} 
 \eta =   \frac{1}{s_{11}},
\quad 
r_i  =  -\frac{s_{1i}}{s_{11}},
\quad
n_{ij} =   \frac{1}{s_{11}}\begin{vmatrix}
                                         s_{11} & s_{1j} \\
                                         s_{1i}  & s_{ij}
                              \end{vmatrix}. 
\end{equation}
Note that in passing from the lower half-space to the upper 
half-space, only $r_6$ and $n_{26}$ change signs.
 
Ting (2003) proved that for any positive or negative integer 
$n$, the matrix $\mathbf{N}^n$ has the structure,
\begin{equation} 
 \mathbf{N}^n
     =  \begin{bmatrix}
          \mathbf{N_1^{(n)}} & \mathbf{N_2^{(n)}} \\
          \mathbf{K^{(n)}}   & \mathbf{N_1}^\text{(\textbf{n})T}  
        \end{bmatrix}, 
\quad \text{with} \quad 
\mathbf{K^{(n)}} = \mathbf{K}^\text{(\textbf{n})T}, \quad 
\mathbf{N_2^{(n)}} = \mathbf{N_2}^\text{(\textbf{n})T}, 
\end{equation}
see also Currie (1979). 
Hence the following matrix $\mathbf{M^{(n)}}$ is symmetric,
\begin{equation} 
\mathbf{M^{(n)}} \equiv  \widehat{\mathbf{I}}\mathbf{N}^n, 
\quad 
 \widehat{\mathbf{I}} = 
  \begin{bmatrix} 0 & \mathbf{1} \\ \mathbf{1} & 0 \end{bmatrix}, 
\quad 
\mathbf{1} = 
  \begin{bmatrix} 1 & 0 \\ 0 & 1 \end{bmatrix}.
\end{equation}
Now premultiply both sides of \eqref{EqnMotion}$_1$ by 
$-i \overline{\mbox{\boldmath $\xi$}}' 
  \widehat{\mathbf{I}}\mathbf{N}^{n-1}$ and add the complex conjugate 
quantity to obtain 
$\overline{\mbox{\boldmath $\xi$}}' \cdot 
   \mathbf{M^{(n)}}\mbox{\boldmath $\xi$} + 
 \overline{\mbox{\boldmath $\xi$}} \cdot 
   \mathbf{M^{(n)}}\mbox{\boldmath $\xi$}' = 0$. 
Assuming that the IAWs vanish at great distance from the interface 
($\mbox{\boldmath $\xi$}(\infty) = \mathbf{0}$), integration 
between $x_2=0$ and $x_2 = \infty$ yields, 
\begin{equation} \label{equations}
\overline{\mbox{\boldmath $\xi$}}(0)  \cdot 
   \mathbf{M^{(n)}}\mbox{\boldmath $\xi$}(0) = 
\overline{\mbox{\boldmath $\xi$}}(0)  \cdot 
   \widehat{\mathbf{I}}\mathbf{N}^n \mbox{\boldmath $\xi$}(0) =  0.
\end{equation} 
This relationship is valid for any type of anisotropy. 
Because of the Cayley-Hamilton theorem, it yields five linearly 
independent equations for a $6 \times 6$ matrix $\mathbf{N}$, 
and three equations for a $4 \times 4$ matrix $\mathbf{N}$, as here. 
Note that for surface wave boundary conditions, the tractions are zero 
at $x_2 = 0$, and the relationship reduces to (Taziev, 1989): 
$\overline{\mathbf{u}}(0)  \cdot \mathbf{K^{(n)}}\mathbf{u}(0) = 0$. 
At $n = 1, 2, -1$, the following expressions are found for 
$\mathbf{M^{(n)}}$,
\begin{equation}
\mathbf{M^{(1)}}=
 \begin{bmatrix}
 X - \eta  &       &          &         \\ 
 0         &  X    &          &         \\ 
 -r_6      &  -1   & n_{66}   &         \\ 
 -r_2      &   0   & n_{26}   &  n_{22}
 \end{bmatrix},
\end{equation}
\begin{equation}
\mathbf{M^{(2)}}=
 \begin{bmatrix}
 2r_6(\eta-X)            &            &            &           \\  
  \eta-(1+r_2)X          &    0       &            &           \\  
r_2+r^2_6-n_{66}(\eta-X) & r_6+n_{26}X&-2(n_{26}+r_6n_{66})&   \\ 
r_2r_6-n_{26}(\eta-X)    & r_2+n_{22}X&-n_{22}-r_2n_{66}-r_6n_{26} 
                                                   &  -2r_2n_{26}
\end{bmatrix},
\end{equation}
and
\begin{equation}
\mathbf{M^{(-1)}}=
 \frac{1}{\Delta}
 \begin{bmatrix} 
            d            &                     &        &         \\ 
X[r_2r_6+n_{26}(\eta-X)] &     a               &        &         \\ 
X[r_6 n_{22}-r_2n_{26})] &     b               &    c   &         \\ 
            e            &r_2r_6+n_{26}(\eta-X)& r_6n_{22}-r_2n_{26} 
                                                        &    f
\end{bmatrix},
\end{equation}
with $\Delta = \text{det } \mathbf{N}$ and 
\begin{align} 
& a = (1 - n_{66} X)(\eta-X) - r_6^2 X , \quad 
  b = -r_2 + X(r_2 n_{66} - r_6 n_{26}), 
\nonumber \\ 
& c = -n_{22}+ X(n_{22}n_{66}- n_{26}^2),
\nonumber \\ 
& d = -X[r_2^2 + n_{22}(\eta-X)], \quad \quad \quad 
  e = -r_2^2 - n_{22}(\eta-X), 
\nonumber \\ 
& f = 2r_2r_6n_{26} - r_2^2n_{66} - r_6^2n_{22}
        - (\eta-X)(n_{22}n_{66} - n_{26}^2). 
\end{align} 
 
Corresponding to the first effective boundary condition \eqref{BC}$_1$ 
is the interface acoustic mode IAW1, for which the quadrivector 
$\mbox{\boldmath $\xi$}(0)$ is in the form,
\begin{equation}
\mbox{\boldmath $\xi$}(0) = U_2(0)[0, 1, \alpha, 0]^\text{T},
\end{equation} 
(say), so that equations \eqref{equations} read
\begin{equation} 
M^{(n)}_{22} + M^{(n)}_{23} (\alpha + \overline{\alpha})
     + M^{(n)}_{33} \alpha \overline{\alpha} = 0.
\end{equation}
Choosing in turn $n = -1, 1, 2,$ an homogeneous linear system of 
equations follows,
\begin{equation} \label{secul1}
\begin{bmatrix} 
   a  &    b        &   c                   \\
   X  &    -1       &  n_{66}               \\
   0  & r_6+n_{26}X & -2(n_{26}+r_6n_{66}) 
\end{bmatrix}
\begin{bmatrix} 
 1 \\ \alpha +\overline{\alpha} \\ \alpha\overline{\alpha}
\end{bmatrix} 
  = \mathbf{0}.
\end{equation}

Similarly, corresponding to the second effective boundary condition 
\eqref{BC}$_2$ is the interface acoustic mode IAW2, for which the 
quadrivector $\mbox{\boldmath $\xi$}(0)$ is in the form, 
\begin{equation}
\mbox{\boldmath $\xi$}(0) = U_1(0)[1, 0, 0, \alpha]^\text{T},
\end{equation} 
(say), so that equations \eqref{equations} read
\begin{equation} 
M^{(n)}_{11} + M^{(n)}_{14} (\alpha + \overline{\alpha})
     + M^{(n)}_{44} \alpha \overline{\alpha} = 0.
\end{equation}
Choosing in turn $n = -1, 1, 2,$ an homogeneous linear system of 
equations follows,
\begin{equation} \label{secul2}  
\begin{bmatrix} 
   d          &    e                &   f        \\
 \eta -X      &  r_2                &  -n_{22}   \\
2r_6(\eta-X)  &r_2r_6-n_{26}(\eta-X)&-2r_2n_{26} 
\end{bmatrix}
\begin{bmatrix} 
 1 \\ \alpha +\overline{\alpha} \\ \alpha\overline{\alpha}
\end{bmatrix} 
  = \mathbf{0}.
\end{equation}

The secular equations for IAW1 and IAW2 correspond to the vanishing 
of the determinant of the $3 \times 3$ matrices given in 
\eqref{secul1} and \eqref{secul2}, respectively.
They are both cubics in $X = \rho v^2$. 
For purposes of comparison, recall that a homogeneous bulk wave 
propagates in the $Ox_1$ direction at a speed $v_a$ such that 
 $X = \rho v_a^2$ is root of the quadratic, 
\begin{equation} \label{seculBulk}
\Delta = \text{det } \mathbf{N} = 0, 
\end{equation} 
and that a surface Rayleigh wave propagates over either half-space 
along $Ox_1$ at a speed $v_R$ such that $X = \rho v_R^2$ is root of 
the quartic (Currie, 1979; Destrade, 2001; Ting, 2002), 
\begin{equation}  \label{seculSAW}
\begin{vmatrix} 
      d      &    a   & X[r_2r_6 + n_{26}(\eta-X)] \\
   X-\eta    &    X   &         0                  \\
2r_6(\eta-X) &    0   & \eta - (1+r_2)X 
\end{vmatrix} = 0.
\end{equation}

Figure 3 shows the influence of the twinning angle $\theta$ upon 
the wave speeds,
for (a) Gallium Arsenide and (b) Silicon.
The respective elastic stiffnesses ($10^{10}$ N/m$^2$) and mass 
densities (kg/m$^3$) are (Royer \& Dieulesaint, 1996): 
$c_{11} = 11.88$, $c_{12} = 5.38$,  $c_{66} = 5.94$, $\rho = 5307$, 
and  $c_{11} = 16.56$, $c_{12} = 6.39$,  $c_{66} = 7.95$, 
$\rho = 2329$.
Similar comments apply to both crystals. 
The secular equation \eqref{secul2} corresponds to a non-physical 
leaky wave IAW2 with a speed (upper full curve) which is always 
supersonic (always above the  curve for the speed given by 
\eqref{seculBulk} of a bulk shear wave (upper dotted curve)).
The secular equation \eqref{secul1} corresponds to a Stoneley 
wave IAW1 with a speed (lower full curve) which is always larger 
than the speed of a Rayleigh wave propagating in either half-space 
(always above the  curve for the speed given by \eqref{seculSAW} 
(lower dotted curve)).
\begin{figure}
\centering
\mbox{\subfigure[Gallium Arsenide]{\epsfig{figure=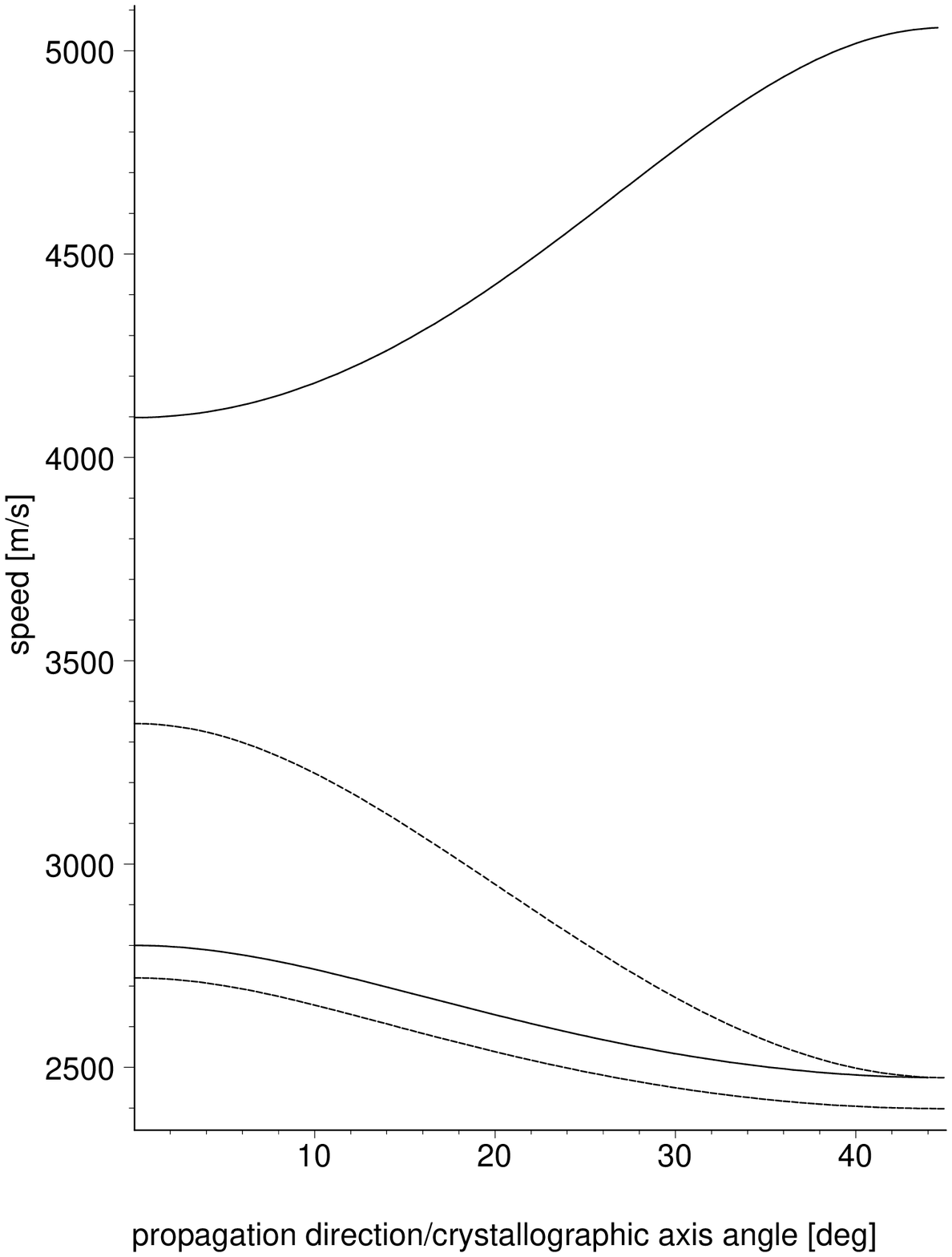,
width=.4\textwidth}}
  \quad \quad
     \subfigure[Silicon]{\epsfig{figure=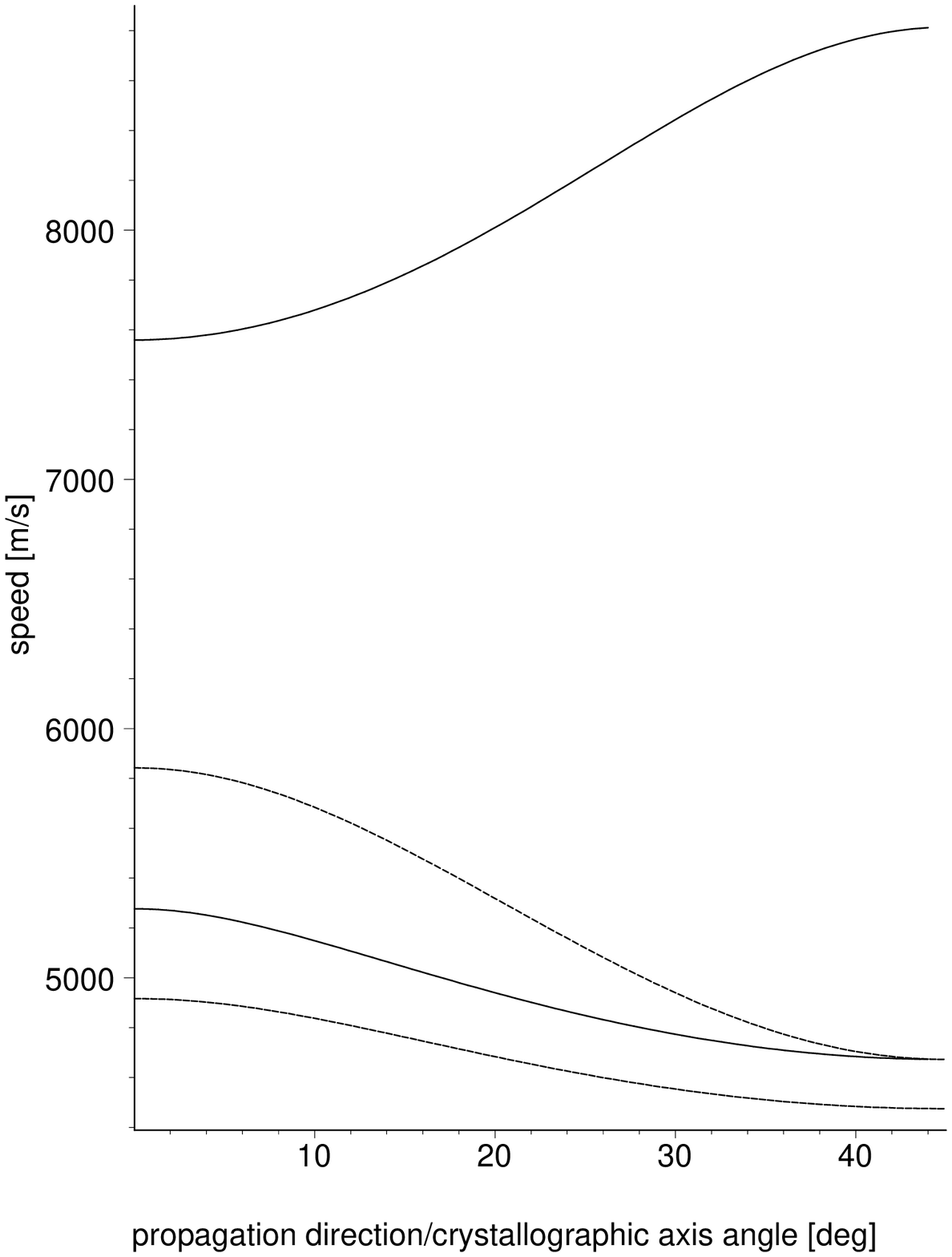,
width=.4\textwidth}}}
\caption{Interface wave speeds in twinned crystals: 
leaky wave (upper solid) and Stoneley wave (lower solid); 
also included: quasi bulk shear wave (upper dashed) and Rayleigh wave 
(lower dashed).}
\end{figure}

\appendix 
 \renewcommand{\thesection}{\Alph{section}}

\section*{Appendix} 
\setcounter{section}{1} 
\setcounter{equation}{0} 
        
Here the effective boundary conditions \eqref{BC} 
for twinned crystals are derived.

Solutions to the equations of motion \eqref{EqnMotion} are in the form 
$\mbox{\boldmath $\xi$}(kx_2) = \mbox{\boldmath $\xi_0$} e^{ipkx_2}$. 
By substitution, $\mbox{\boldmath $\xi_0$}$ is found from the adjoint 
matrix to $\mathbf{N} - p \mathbf{1}$ as
\begin{equation}
\mbox{\boldmath $\xi_o^i$} = [a_i, b_i, c_i, d_i]^\text{T},
\end{equation}
with
\begin{align}
& a_i = [(1+r_2)X-\eta]p_i + r_2r_6X + n_{26}(\eta-X)X, 
\nonumber \\
& b_i = -Xp_i^2 - 2r_6Xp_i - [1+r_6^2+n_{66}(\eta-X)]X + \eta,
\nonumber \\
& c_i = p_i^2 + (r_6-n_{26}X)p_i - r_2(1-n_{66}X) - r_6n_{26}X,
\nonumber \\
& d_i = -p_i^3 - 2r_6p_i^2 + [r_2 - r_6^2 - n_{66}(\eta-X)]p_i
 + n_{26}(\eta-X) + r_2r_6.
\end{align}
Here $p$ is a root of the characteristic polynomial 
$\text{det }(\mathbf{N} - p \mathbf{1}) = 0$, which is given 
explicitly by \eqref{quartic-p}.
For the lower half-space, only the roots $p_1$, $p_2$ with positive 
imaginary parts are kept in order to ensure decay away from the 
interface $x_2=0$.
Thus the solution for $x_2>0$ is 
\begin{equation}
\mbox{\boldmath $\xi$} (kx_2) = 
 \beta_1 \mbox{\boldmath $\xi^1_0$}e^{ip_1kx_2}
  + \beta_2 \mbox{\boldmath $\xi^2_0$}e^{ip_2kx_2},
\end{equation}
for some constants $\beta_1$ and $\beta_2$.

In the upper half-space $x_2<0$, the $p$'s are changed to their 
opposite so that the roots of the characteristic polynomial with 
negative imaginary parts are $-p_1$, $-p_2$.
Moreover, $r_6$ and $n_{26}$ also change signs, so that the 
eigenvectors  $\mbox{\boldmath $\widehat{\xi}_0$}$ are in the form,
\begin{equation}
\mbox{\boldmath $\widehat{\xi}_o^i$} = [-a_i, b_i, c_i, -d_i]^\text{T},
\end{equation}
and the solution for $x_2<0$ is 
\begin{equation}
\mbox{\boldmath $\widehat{\xi}$} (kx_2) = 
 \widehat{\beta}_1 \mbox{\boldmath $\widehat{\xi}^1_0$}e^{-ip_1kx_2}
 +\widehat{\beta}_2 \mbox{\boldmath $\widehat{\xi}^2_0$}e^{-ip_2kx_2},
\end{equation}
for some constants $\widehat{\beta}_1$ and $\widehat{\beta}_2$.

It follows that the continuity of the displacements and of the 
tractions across the interface $x_2=0$ reads 
\begin{equation}
 \beta_1 \mbox{\boldmath $\xi^1_0$}
  + \beta_2 \mbox{\boldmath $\xi^2_0$} = 
 \widehat{\beta}_1 \mbox{\boldmath $\widehat{\xi}^1_0$}
 +\widehat{\beta}_2 \mbox{\boldmath $\widehat{\xi}^2_0$}.
\end{equation}
This system of equations is re-written as 
\begin{equation} \label{system}
\begin{bmatrix}
 \mathbf{A} & \mathbf{A} \\
 \mathbf{B} & -\mathbf{B}
\end{bmatrix} \begin{bmatrix}
               \mbox{\boldmath $\beta$} \\ 
               \mbox{\boldmath $\widehat{\beta}$}
              \end{bmatrix} = \mathbf{0},
\end{equation}
where 
\begin{equation} 
\mathbf{A} = \begin{bmatrix}
              a_1 & a_2 \\
              d_1 & d_2
\end{bmatrix},
\quad 
\mathbf{B} = \begin{bmatrix}
              b_1 & b_2 \\
              c_1 & c_2
\end{bmatrix},
\quad 
 \mbox{\boldmath $\beta$} =  \begin{bmatrix}
                               \beta_1 \\  \beta_2
                             \end{bmatrix}
\quad 
\mbox{\boldmath $\widehat{\beta}$}
       =  \begin{bmatrix}
            \widehat{\beta}_1 \\  \widehat{\beta}_2
          \end{bmatrix}.
\end{equation}
The determinant of the $4 \times 4$ matrix in \eqref{system} is 
$4 \text{ det } \mathbf{A} \text{ det } \mathbf{B}$.
It is zero when either (a) $\text{det } \mathbf{A} = 0$, and then 
$\mathbf{B}  \mbox{\boldmath $\beta$}
 - \mathbf{B} \mbox{\boldmath $\widehat{\beta}$} = \mathbf{0}$ 
yields $\mbox{\boldmath $\widehat{\beta}$}=\mbox{\boldmath $\beta$}$;
or (b) $\text{det } \mathbf{B} = 0$, and then 
$\mathbf{A}  \mbox{\boldmath $\beta$}
 + \mathbf{A} \mbox{\boldmath $\widehat{\beta}$} = \mathbf{0}$ yields 
$\mbox{\boldmath $\widehat{\beta}$}= - \mbox{\boldmath $\beta$}$.
In case (a), $\mathbf{A}  \mbox{\boldmath $\beta$}
 + \mathbf{A} \mbox{\boldmath $\widehat{\beta}$} =  \mathbf{0}$ means 
$\mathbf{A}  \mbox{\boldmath $\beta$}=  \mathbf{0}$ , that is 
\begin{equation}
u_1(0) = t_{22}(0) = 0;
\end{equation}
in case (b), $\mathbf{B}  \mbox{\boldmath $\beta$}
 - \mathbf{B} \mbox{\boldmath $\widehat{\beta}$} =  \mathbf{0}$ means 
$\mathbf{B}  \mbox{\boldmath $\beta$}=  \mathbf{0}$ , that is 
\begin{equation}
u_2(0) = t_{12}(0) = 0.
\end{equation}


        
\newpage

\bigskip

\noindent

{\Large\textbf{List of Figures.}}

\bigskip
\noindent
\textbf{Figure 1: Twinned crystal.}

\bigskip
\noindent
\textbf{Figure 2: Cutting, rotating, and bonding of a rhombic crystal.}

\bigskip
\noindent
\textbf{Figure 3: Interface wave speeds in twinned crystals: 
leaky wave (upper solid) and Stoneley wave (lower solid); 
also included: quasi bulk shear wave (upper dashed) and Rayleigh wave 
(lower dashed).}

\medskip

Figure 3(a): Gallium Arsenide.

\noindent
Legend on graduated horizontal axis: ``propagation direction/crystallographic axis angle [deg].''

\noindent
Legend on graduated vertical axis: ``speed [m/s].''

\medskip

Figure 3(b): Silicon.

\noindent
Legend on graduated horizontal axis: ``propagation direction/crystallographic axis angle [deg].''

\noindent
Legend on graduated vertical axis: ``speed [m/s].''.


\end{document}